\documentclass[aps, twocolumn, prx, superscriptaddress]{revtex4-2}

\usepackage{amsmath}
\usepackage{amsfonts}
\usepackage{amssymb}
\usepackage{amstext}
\usepackage{amsthm}
\usepackage{physics}
\usepackage{braket}
\usepackage{dsfont}

\usepackage[T1]{fontenc}

\usepackage{graphicx}
\usepackage[dvipsnames]{xcolor}
\usepackage[notransparent]{svg}
\usepackage{tikz}

\usepackage{booktabs}

\definecolor{myRed}{HTML}{de3f2b}
\definecolor{myYellow}{HTML}{d8a26f}
\usepackage[
  colorlinks=true,
  linkcolor=myYellow,
  citecolor=BrickRed,
  filecolor=BrickRed,
  urlcolor=BrickRed
]{hyperref}

\usepackage[capitalize]{cleveref}

\newcommand{\qmaddress}{\affiliation{Quantum Motion, 9 Sterling Way, London N7 9HJ, United Kingdom}}
\newcommand{\konstanzadress}{\affiliation{Department of Physics and IQST, University of Konstanz, 78457 Konstanz, Germany}}
\newcommand{\ucladress}{\affiliation{Department of Physics and Astronomy \& London Centre for Nanotechnology, University College London, 17-19 Gordon Street, London WC1E 6BT, United Kingdom}}

\newcommand{\skipLine}{\nonumber \\}

\begin{document}

\title{Electron shuttling as a probe for charge defects}
\author{Hamza Jnane}
\thanks{hamza@quantummotion.com}
\qmaddress
\author{Tamas Daniel Lipovics}
\thanks{daniel.lipovics.20@ucl.ac.uk}
\qmaddress
\ucladress
\author{Guido Burkard}
\thanks{guido.burkard@uni-konstanz.de}
\konstanzadress

\begin{abstract}
    Silicon spin qubits are a leading platform for scalable quantum computing, but their performance is limited by charge noise, widely attributed to two-level fluctuators (TLFs). The location of individual TLFs is generally unknown, and existing methods to localise them does not scale favourably with device size. Here, we show that electron shuttling turns a single mobile spin into a scanning probe of individual defects. We show that by shuttling a spin over a range of distances and tracking its coherence loss, one can localise defects along the channel and constrain their switching rate and fluctuation amplitude. Because one shuttled electron sweeps an extended region, the approach scales more favourably than previous methods. Our protocol requires no additional hardware and provides a practical route to mapping the charge-defect landscape of large silicon devices, enabling defect-aware calibration and avoidance in shuttling-based architectures.
\end{abstract}
\maketitle      
\textbf{\emph{Introduction ---}}
Silicon spin qubits are strong candidates for building large-scale quantum computers due to their small footprint, fast gate operations and compatibility with industry standards \cite{Maurand2016AQubit,gonzalez_zalba_scaling_2021,zwerver_qubits_made_2022,chittockwood2024exchangecontrolmosdouble, steinacker_industry-compatible_2025, George2025, burkard_2023}. Recent experiments have demonstrated single- and two-qubit gate fidelities exceeding 99.99\% \cite{wu_five_nines_2025} and 99.5\% \cite{mills_2022, noiri_2022,xue_2022} respectively, alongside high-fidelity readout ($>99\%$) \cite{mills_2022,constance_readout_2025, steinacker_industry-compatible_2025, takeda_rapid_2024} and shuttling \cite{Yoneda_2021, Seidler_2022, de_smet_shuttling_2024}. The latter capability being a key component of future fault-tolerant architectures \cite{xue_qubus_2024, adam_two_by_N_2024, siegel_snakes_2025, li_crossbar_2018, Cai_2023, jnane_global_2025}. However,  these fidelities were achieved in small-scale devices, and to maintain them at scale, environmental noise inherent to solid-state platforms must be mitigated. While nuclear spin noise can be reduced through isotopic purification \cite{nuclearspin1, nuclearspin2}, charge noise remains the dominant source of errors \cite{burkard_2023}. 

Charge noise is a ubiquitous challenge in solid-state systems \cite{weissman_1998}. It arises from stochastic fluctuations of local electric fields caused by the charging and discharging of traps, or rearrangements of defects in the lattice. Noise in amorphous materials is traditionally modelled by ensembles of two-level fluctuators (TLFs) \cite{Anderson01011972, Phillips1972TunnelingSI}. Each TLF is characterised by its position in space, its spatial fluctuation amplitude (measured through voltage shifts at nearby electrodes) and its switching rate \cite{ye2024characterization}. 
Candidate microscopic origins include the
$P_b$ centre at the Si/SiO$_2$ interface and the E' centre in bulk SiO$_2$ \cite{defect_candidates}.
Because of their inherently random nature, these fluctuations constitute a source of decoherence which leads to a degradation in gate fidelity \cite{jnane_ab_initio_exchange_2024, cifuentes_path_integral_exchange_2023, sheahta_modelling_2023}. If one could characterise TLFs, error mitigation techniques \cite{cai_qem_2023, jeon_qec_qem_2026} could, e.g., be leveraged to improve the device's performance. In some architectures, one could even imagine avoiding zones of the device where TLFs are present to remove the possibility of logical errors \cite{siegel_snakes_2025}.

Traditionally, charge noise is measured via the Coulomb-blockade flank method \cite{Connors2019}, dynamically-decoupled exchange oscillations \cite{Connors2022}, or RF-reflectometry of a single-electron transistor \cite{rf_reflectometry}. These extract a noise power spectral density or a superposition of random telegraph signals \cite{sparsebath}, quantifying the local electric-field strength but revealing little about the atomistic
origin of the noise, and none can pinpoint the location of an individual TLF.

Several techniques instead target defect location. Electron spin resonance (ESR) enabled the original discovery of $P_b$ centres \cite{Nishi_1972}, but yields only an aggregate signal, requires invasive etching or sample reorientation to recover spatial information. Moreover, it cannot be used during QPU operation. Scanning tunnelling microscopy \cite{STM} and tip-induced mobile quantum dots \cite{cakar2024} map defect positions directly, but only on an open device face. However, the defects we are interested in sit at a buried interface or in the bulk. Most recently, Rojas-Arias \textit{et al.} \cite{rojas_charge_noise_2026} located charge defects from noise cross-correlations between multiple quantum dots, but the number of measurements required scales with device size.

Here, we propose to use a single mobile quantum dot to probe charge defects across large sections of a device. By leveraging electron shuttling, we show that one can obtain properties of TLFs along the shuttling path through a simple experiment. More precisely, we demonstrate that measuring the average phase a spin qubit accumulates while being shuttled back and forth for various distances allows for the characterisation of the position of charge defects and of its switching rate. Conveyor-belt electron shuttling has previously been used to map the valley splitting  \cite{volmer2024}, as well as the valley phase and g-factor \cite{volmer2026}.

\begin{figure*}
    \centering
    \includegraphics[width=.9\linewidth]{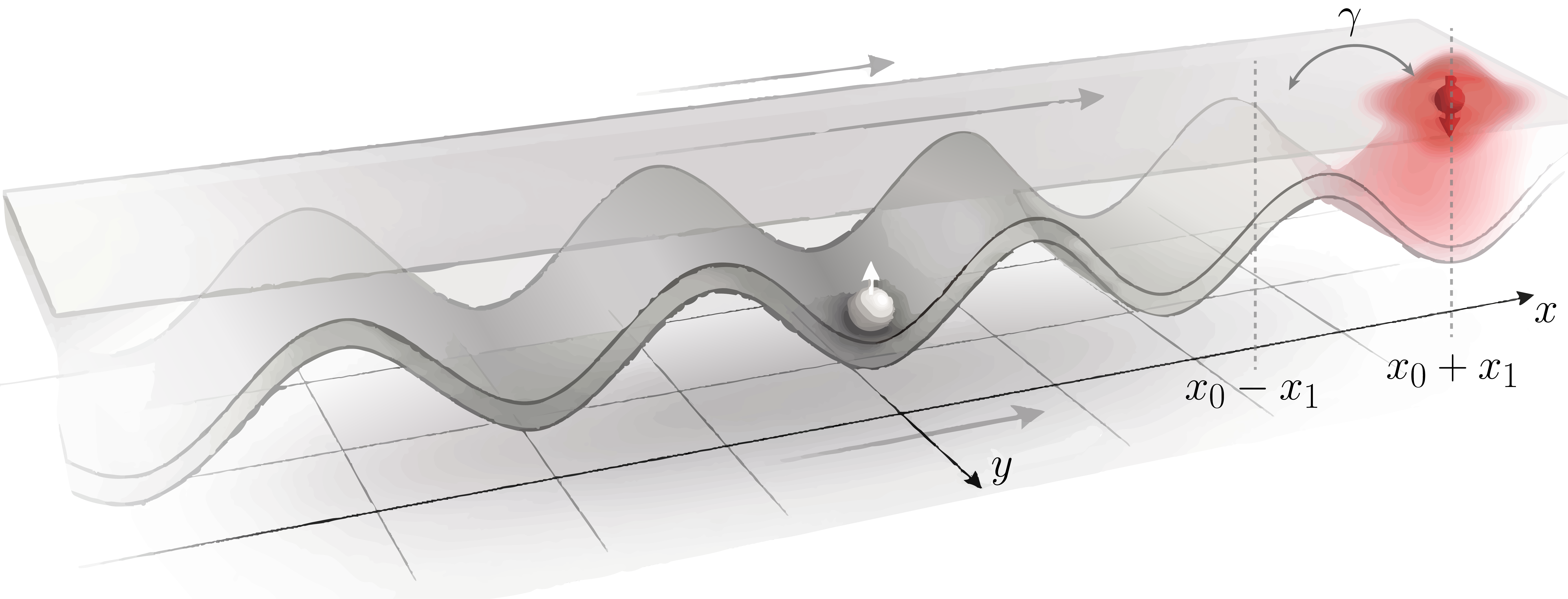}
    \caption{Illustration of shuttling-based defect probing. An electron is shuttled back and forth in the $x$-direction. For simplicity, we represent a charge defect fluctuating in the $x$-direction in red but consider fluctuations in 3D in the main text. The closer the shuttled electron is to the defect (red zone in the potential landscape), the more spin dephasing it experiences. The goal is to leverage this to characterise the average position $x_0$ of the TLF by measuring the average phase the electron has accumulated during its trip. 
    This illustration is inspired by Fig.~2 of \cite{burkard_2023}.}
    \label{fig:illustration_protocol}
\end{figure*}

\textbf{\emph{Preliminaries ---}} \emph{Intuition.}
Since spin-orbit coupling makes the electron $g$-factor sensitive to the local electric field, a spin qubit shuttled past a charge defect (see \cref{fig:illustration_protocol}) dephases more the closer it passes to it. Shuttling back and forth over increasing distances and tracking the resulting dephasing should therefore reveal a kink at the defect's position. We formalise this below.

\emph{The model.}
For simplicity, we focus on the case of a single charge defect placed at the Si/SiO$_2$ interface and suppose that its position evolves in time as $\mathbf{r}_{d}(t) =(x_d(\tau), y_d(\tau),z_d(\tau))= \mathbf{r}_0 + \boldsymbol{\delta r}(t)$ with the random fluctuations $\boldsymbol{\delta r}(t)$ switching between $\pm \mathbf{r}_1$ at a rate $\gamma$, following a Poisson distribution. We denote by $\mathbf{r}_{d}^{\pm} = \mathbf{r}_0 \pm \mathbf{r}_1$ the two positions of the charge. The impact of the TLF is modelled by a modification of the trapped electron's g-factor, which is proportional to the electric field it generates (see \cref{app:g_factor} for more details),
\begin{align}
    g_{\rm TLF}(t) = \sigma\frac{z_d(t)}{|\mathbf{r}(t)-\mathbf{r}_{d}(t)|^3}
\end{align}
where $\mathbf{r}(t)$ is the position of the shuttled electron and $\sigma$ (defined in \cref{app:g_factor}) can be understood as an effective cross section: the smaller $\sigma$ is, the closer a TLF needs to be to significantly impact the qubit.

The Hamiltonian of the system reads, 
\begin{align}
    H(t) = \frac{1}{2}\mu_B\big(g_0+g_{\rm int}(t)+g_{\rm TLF}(t)\big)B_0\sigma_z,
\end{align}
where $g_0$ is the bulk g-factor, $g_{\rm int}$ the additional contribution due to the rough interface which evolves in time in the moving electron's rest frame, $B_0$ is the Zeeman splitting field, and $\sigma_z$ is the Pauli Z operator.
We first focus on the impact of the TLF, which should be dominant. We thus assume that the interface is flat, yielding $g_{\rm int}(t) = 0$ and
\begin{align}
    H(t) = \frac{1}{2}\mu_B\left(g_{0}+\sigma\frac{z_d(t)}{|\mathbf{r}(t)-\mathbf{r}_{d}(t)|^3}\right)B_0\sigma_z.
    \label{eq:hamil}
\end{align}

\emph{The protocol.}
Let's now consider the evolution of the state of the electron spin qubit in time under the Hamiltonian \cref{eq:hamil}. At time $t$, its density matrix reads
\begin{align}
    \rho(t) = \begin{pmatrix}
            \rho_{00}(0) & \rho_{01}(0)\exp(-i\phi(t)) \\
             \rho_{10}(0)\exp(i\phi(t)) & \rho_{11}(0)\\
    \end{pmatrix}.
\end{align}

The accumulated phase $\phi(t)$ is given by, 
\begin{align}
    \phi(t) &= \phi_0(t) + \phi_1(t) \skipLine 
    &= \frac{\mu_Bg_0}{\hbar}B_0t  + \sigma^{'}\int_0^t\frac{z_d(\tau)}{|\mathbf{r}(\tau)-\mathbf{r}_{d}(\tau)|^3}d\tau.
\end{align}
where $\sigma^{'} = \mu_B \sigma B_0/\hbar$. As the TLF fluctuates randomly over time, the quantity of interest is the average phase factor obtained after multiple experiments (i.e., multiple realisations of the TLF signal) which we denote by $W(t)$ \cite{mokeev_decoherence_2024}. $W(t)$ is a dephasing factor, ranging from $|W(t)| = 0$ (fully decohered) to $|W(t)| = 1$ (no dephasing). More formally, the dephasing factor is defined as 
\begin{align}
    W(t) = \mathbb{E}\left[ \exp\left( -i\phi(t)\right)\right],
\end{align}
where the average is performed over the TLF's distribution.
From now on, we discard the phase $\phi_0$ (as it is independent of the presence and location of any defect and can be calibrated away) and focus on $\phi_1$ as it contains the impact of the TLF. 

Our sensing protocol is as follows: shuttle an electron forward for a time $T/2$ and a distance $L$ and backward for the rest of time ($T/2$) at a constant speed $v$. For simplicity, we will focus on an electron shuttled along a 1D line in the $x$-direction, $\mathbf{r}(t)=(vt,0,0)$, as illustrated in \cref{fig:illustration_protocol}. We name this protocol forward-backward (FB). 
We show below that by plotting the evolution of the dephasing factor with $L=vT/2$ for the FB protocol, we can recover properties of the TLF. 

\textbf{\emph{Results ---}} \emph{Analytical derivations.}
The phase accumulated over this experiment can be decomposed as follows,
\begin{align}
    \phi_1(T) = \sigma^{'}\Bigg(\int_0^{T/2}\!\!\!\frac{z_d(\tau)}{((v\tau-x_d(\tau))^2 + y_d(\tau)^2 + z_d(\tau)^2)^{3/2}} d\tau \nonumber \\
    + \int_{T/2}^{T}\frac{z_d(\tau)}{((2L-v\tau-x_d(\tau))^2 + y_d(\tau)^2 + z_d(\tau)^2)^{3/2}} d\tau\Bigg),
\end{align}
with $2L$ being the distance the electron has travelled.
Let $w_{\pm}(t)$ be the average accumulated phase when the position of the TLF is fixed to $\mathbf{r}_{d}^{\pm}$; then, we have
\begin{align}
    w_\pm(t) = \mathbb{E}\Big[ \exp\left(-i\phi_1(t) \right)\delta_{\mathbf{r}_d(t), \mathbf{r}_{d}^\pm}\Big],
\end{align}
where $\delta_{\mathbf{r}_d(t), \mathbf{r}_{d}^\pm}$ is a Kronecker delta function, i.e., $\delta_{\mathbf{r}_d(t), \mathbf{r}_{d}^\pm} = 1$ when $\mathbf{r}_d(t) = \mathbf{r}_{d}^\pm$ and $\delta_{\mathbf{r}_d(t), \mathbf{r}_{d}^\pm} = 0$ otherwise. Note that we thus have $W(t) = w_+(t) + w_-(t)$. 
We show in \cref{app:proofs} that the vector $\mathbf{w}(t) = (w_+(t),w_-(t))^T$ is a solution of the system of differential equations, 
\begin{align}
    \frac{d\mathbf{w}}{dt}(t) = M(t)\mathbf{w}(t),
    \label{eq:sys_diff_eq}
\end{align}
where
$M(t) = \begin{pmatrix}
    -if_+(t)-\gamma & \gamma \\
    \gamma & -if_-(t)-\gamma
\end{pmatrix}$, and,
\begin{align}
    &f_\pm(t)  \skipLine
    &= 
    \begin{cases}
        \frac{z_d^\pm\sigma^{'}}{\left(\left(vt-x_d^\pm\right)^2 + \left(y_d^\pm\right)^2 + \left(z_d^\pm\right)^2\right)^{3/2}}, & \text{for } 0\leq t\leq T/2,\\
        \frac{z_d^\pm\sigma^{'}}{\left(\left(2L-vt-x_d^\pm\right)^2 + \left(y_d^\pm\right)^2 + \left(z_d^\pm\right)^2\right)^{3/2}}, & \text{for } T/2\leq t\leq T.\\
    \end{cases}
\end{align}
Our Eq.~\eqref{eq:sys_diff_eq} can be viewed as an extension of the case of a stationary qubit coupled to a single TLF \cite{bergli2009}. Note that we explore alternative schemes in \cref{app:alt_protocols}, namely forward-wait-backward (FWB) and forward-wait-forward (FWF).

While the matrix $M$ does not commute with itself at different times and as such we cannot obtain a simple closed form of the solution to Eq.~\eqref{eq:sys_diff_eq}, we can explore limiting cases analytically. Specifically, we explore below the $\gamma T \ll 1$ and $\gamma T \gg 1$ limits before exploring a more realistic scenario by solving \cref{eq:sys_diff_eq} numerically.

\emph{Static defect ($\gamma T \ll 1$).}
In the absence of fluctuations, one finds
\begin{equation}
\begin{aligned}
    \frac{dw_\pm}{dt}(t) &= -if_\pm(t)w_\pm(t), 
\end{aligned}
\end{equation}
which leads to the solutions 
\begin{equation}
\begin{aligned}
    w_\pm(t) &= w_\pm(0)\exp\left(-i\int_0^tf_\pm(\tau)d\tau\right). 
\end{aligned}
\end{equation}
Without loss of generality, we assume that $w_+(0)=w_-(0) = 1/2$. The defect is static over an experiment, but uniformly fluctuates between its two states from one shuttling round to the next. As explained above, we consider a charge fluctuating symmetrically 
between the two positions $\mathbf{r}_{d}^{\pm} = (x_0 \pm x_1, y_0 \pm y_1, z_0 \pm z_1)$.

\begin{figure*}
    \centering
    \includegraphics[width=\linewidth]{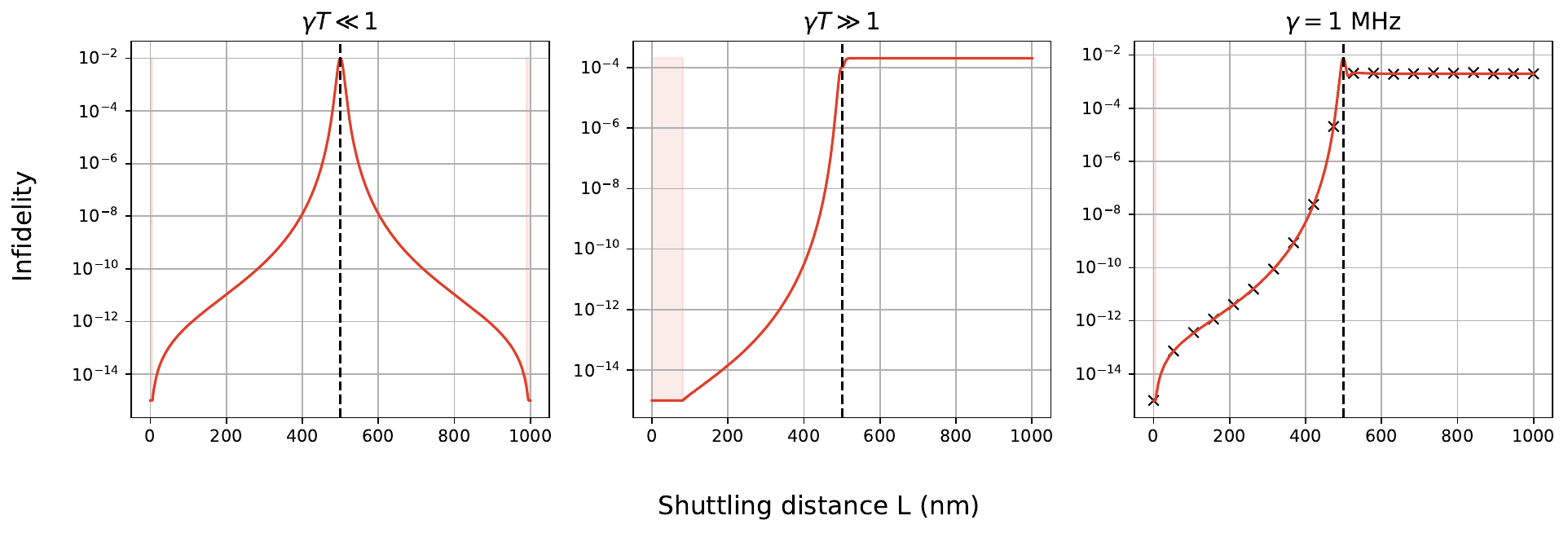}
    \caption{Infidelity ($1-|W|$) in dependence of the shuttling distance $L$ when the TLF is fluctuating in the $x$-direction for the FB protocol (red). We explore three different regimes: static defect ($\gamma T \ll 1$, left), fast fluctuations ($\gamma T \gg 1$, centre) and intermediate fluctuations ($\gamma = 1$ MHz, right). The agreement between the analytically derived results (red line) with Monte-Carlo simulations (crosses) is very good. The mean position of the defect is chosen to be $x_0=500\,{\rm nm}$ (vertical dashed line). Generally, parameters are chosen according to Table~\ref{tab:parameters}, with fluctuations in the $y$- and $z$-direction equal to zero here. Due to numerical instability and precision constraints, values within the red shaded region have been clipped. 
    }
    \label{fig:fluct_gamma}
\end{figure*}

We can now analytically compute the integrals and obtain a closed form of the dephasing factor,
\begin{align}
    W(T) &= \frac{1}{2}\left(e^{-i\Phi_+(L)} + e^{-i\Phi_-(L)}\right) \nonumber \\
    &= e^{-i\frac{\Phi_+(L) + \Phi_-(L)}{2}}\cos\left(\frac{\Phi_+(L) - \Phi_-(L)}{2}\right),
\end{align}
where $\Phi_\pm(L) = \frac{2\sigma^{'}z_d^\pm}{v\beta_\pm^2}\left[ \frac{L-x_d^\pm}{\sqrt{\left(L-x_d^\pm\right)^2+\beta_\pm^2}} + \frac{x_d^\pm}{\sqrt{\left(x_d^\pm\right)^2+\beta_\pm^2}}\right]$ with $\beta_\pm^2=\left(y_d^\pm\right)^2+\left(z_d^\pm\right)^2$. We give more details on the derivation in \cref{app:proofs}.
In \cref{fig:fluct_gamma}, we show the evolution of the \emph{infidelity}, defined as,
\begin{align}
    1-|W(T)| = 1-\left|\cos\left(\frac{\Phi_+(L) - \Phi_-(L)}{2}\right)\right|,
    \label{eq:infidelity_x}
\end{align}
for a charge oscillating in the $x$-direction for realistic parameters as summarised in \cref{tab:parameters}.
\begin{table}[]
    \centering
    \begin{tabular}{l|c}
        \hline
        \hline
         \quad \quad \quad \quad \quad Parameters & Value \\
         \hline
         TLF Average position (nm) & $(x_0,y_0, z_0) = (500, 10, 0.5)$  \\
         TLF Fluctuation amplitude (nm) & $(x_1,y_1, z_1) = (5,5,1)$ \\         
         TLF Switching rate (MHz)  & 1\\
         Shuttling speed (m/s) & $0.1$ \\
         \hline
         \hline
    \end{tabular}
    \caption{Realistic parameters for our protocol. More details on these choices are given in \cref{app:params}.}
    \label{tab:parameters}
\end{table}

We see that we can recover the position of the TLF in the shuttling direction. While the infidelity grows as expected when we get closer to the defect, it decreases after passing it due to destructive interferences between the scenario in which $x_d(0) = x_d^+$ and the one in which $x_d(0) = x_d^-$.

\emph{Fast fluctuations ($\gamma T \gg 1$).}
In this regime, we have a coupled system of differential equations. To simplify the study, we introduce $u_1 = w_+ + w_-$, $u_2 = w_+-w_-$ and rewrite \cref{eq:sys_diff_eq} in terms of these variables,
\begin{equation}
    \begin{aligned}
    \frac{du_1}{dt}(t) &= -i \bar{f}(t)u_1(t) -i\delta f(t)u_2(t), \skipLine
    \frac{du_2}{dt}(t) &= -i \bar{f}(t)u_2(t) -i\delta f(t)u_1(t) - 2\gamma u_2(t),
    \label{eq:sys_diff_eq_2}
\end{aligned}
\end{equation}
with $\bar{f}(t) = 1/2(f_+(t)+f_-(t))$ and $\delta f(t) = 1/2(f_+(t)-f_-(t))$.
We show in \cref{app:proofs} that this leads to an infidelity,
\begin{align}
    1-|W(T)| = 1-\exp\left( -\frac{1}{2\gamma}\int_{0}^{T}\delta{f}(\tau)^2d\tau\right).
\end{align}
In \cref{fig:fluct_gamma}, we plot the evolution of the infidelity for fluctuations in the $x$-direction. Unlike the $\gamma T \ll 1$ case, the infidelity now evolves strictly monotonically with $L$. Intuitively, this is because fast oscillations destroy the interference between the defect's two positions. Nevertheless, we are still able to recover the position of the defect.

\emph{Numerical simulations.}
We numerically solve \cref{eq:sys_diff_eq} for $\gamma = 1$ MHz to study the system outside of the limiting cases. Moreover, we compare our solver's results (red line) with Monte-Carlo simulations of the experiment (crosses) and obtain perfect agreement as seen on the right of \cref{fig:fluct_gamma}. Note that we are still able to detect the position of the TLF for each direction in this more realistic scenario. In this intermediate regime, the infidelity still increases up to defect but drops to a plateau right after. The plateau can be explained by the switching of the TLF destroying the interference between the first and second time the shuttled electrons encounters the defect. Note that by fitting this curve to an experiment one could recover the amplitude and switching rate of the defect.

In \cref{app:other_dir} and \cref{app:more_defects}, we generalise our protocol by exploring the case of a TLF oscillating in other directions and studying the detection of multiple defects respectively.

\begin{figure}
    \centering
    \includegraphics[width=0.5\textwidth]{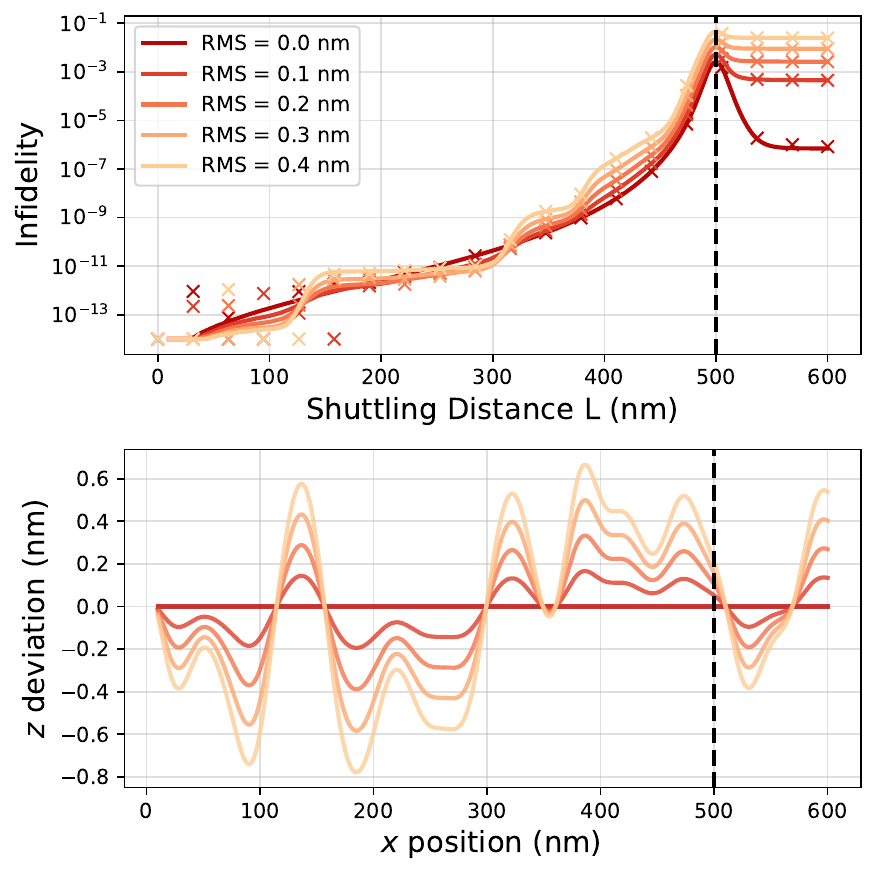}
    \caption{(Top) Evolution of the infidelity ($1-|W|$) as a function of the shuttling distance $L$ along different Si/SiO$_2$ interface (RMS) roughness values. The crosses represent results of Monte-Carlo simulations with $10^5$ shots. (Bottom) Physical interface terrain profiles.}
    \label{fig:roughness}
\end{figure}

\emph{Simulation for rough Si/SiO$_2$ interfaces.} So far we have idealised the shuttling track as a perfectly smooth and sharp interface. To test whether our protocol would succeed for a more realistic interface (\emph{e.g.,} roughness due to thermal oxidation), we treat the qubit conforming rigidly to a rough interface in the $z$-direction, thereby modulating its distance to the defect and the effective g-factor shift. This is justified by the strong confinement fields in $z$ and the large band offset of the Si-SiO$_2$ interface. While it decreases the sensitivity and even introduces qualitative variation unique to surface features, roughness does not fundamentally alter the efficiency of our protocol for realistic RMS values, as seen in \cref{fig:roughness}.

\textbf{\emph{Conclusions ---}}
We tackle the pressing issue of charge noise by presenting a simple electron-shuttling protocol allowing for the detection and localisation of charge defects, and for the extraction of information about its switching rate and fluctuation amplitude. Shuttling a single spin qubit back and forth along a 1D channel past a TLF and measuring its dephasing factor as a function of the shuttled distance reveals these properties directly.

Since the dephasing factor can be measured through a Ramsey-like experiment, our protocol is straightforward to implement. It can be viewed as a simpler version of existing experiments \cite{de_smet_shuttling_2024} requiring no waiting time.
Moreover, depending on the setup, one could try to implement one of the alternative protocols (FWB or FWF) as presented in \cref{app:alt_protocols}.

We believe that our protocol will be helpful for the construction of larger-scale silicon-based quantum computers of different eras. In the near-term, finding out the characteristics of TLFs could be helpful to mitigate their impact through careful pulse design or error mitigation. In the longer-term, one can view our protocol as a calibration step where all charge defects are located and can thus be avoided in shuttling-heavy fault-tolerant architectures \cite{Cai_2023, siegel_snakes_2025}.

There are several future directions for this work. One could envision shuttling other types of qubits, e.g., a singlet-triplet qubit which is more protected against charge noise, and for each distance $L$ pulling the two dots apart so that the qubit becomes more sensitive to the defect. It would also be interesting to leverage more advanced modelling techniques to study the impact of interface roughness, electrode imperfections, and other imperfections of real devices \cite{jack_device_2025}. Finally, our technique could be extended to detecting charge neutral defects as briefly addressed in \cref{app:spin}.

\textbf{\emph{Acknowledgements ---}}
The authors would like to thank Konstantinos Tselios and Thomas Swift who carried out charge noise measurements that informed our simulation parameters. TDL acknowledges funding from the Engineering and Physical Sciences Research Council through the Oxford Quantum Computing and Simulation Hub.

\onecolumngrid
\newpage
\begin{center}
    {\large \textbf{Appendices for ``Electron shuttling as a probe for charge defects''}}\\[0.5em]
\end{center}
\appendix
\crefalias{section}{appendix}
\section{\label{app:g_factor} Impact of a TLF on the g-factor}
In \cref{eq:hamil} we assumed that the $g$-factor was proportional to the field generated by the TLF. Here, we give an analytical formula for the proportionality constant $\sigma$ based on the derivation of Ruskov \emph{et al.} \cite{Ruskov_g_factor_2018}. 
For an in-plane magnetic field with angle $\phi$ and for an applied electric field $F_z$ (confinement in the $z$-direction), Ruskov \emph{et al.} showed that, to first order, the total $g$-factor of an electron in a Si/SiO$_2$ heterostructure is given by,
\begin{align}
    g(\phi, F_z) &= g_0+\delta g(\phi, F_z) = g_0 + \frac{\beta(F_z) \sin(2\phi)-\alpha(F_z)}{\hbar\mu_B}|e|\braket{z}(F_z),
\end{align}
where $\alpha(F_z)$ and $\beta(F_z)$ are the field-dependent Rashba and Dresselhaus spin-orbit coupling terms. For a linear vertical confinement potential we have,
\begin{align}
    \braket{z}(F_z) &= 1.5587\left(\frac{\hbar^2}{2m_l|e|F_z}\right)^{1/3}, \skipLine
    \alpha(F_z) &= \alpha^{'}F_z, \skipLine
    \beta(F_z) &= \beta^{'}F_z,
\end{align}
with $\hbar$ the reduced Planck constant, $m_l$ the longitudinal effective mass, $e$ the electron charge, which leads to,
\begin{align}
    g(\phi, F_z) = g_0 + \delta g(\phi, F_z) = g_0 + 1.5587\frac{|e|^{2/3}}{\mu_B}\left( \frac{1}{2m_l\hbar}\right)^{1/3}\left( \beta^{'}\sin(2\phi)-\alpha^{'}\right) F_z^{2/3},
\end{align}
where $\mu_B$ denotes the Bohr magneton.
We can now compute the variation of the $g$-factor due to a small fluctuation of the electric field,
\begin{align}
    \frac{\partial g}{\partial F_Z}\left(\phi, F_z\right) = 1.5587\frac{2|e|^{2/3}}{3\mu_B}\left( \frac{1}{2m_l\hbar}\right)^{1/3}\left( \beta^{'}\sin(2\phi)-\alpha^{'}\right) F_z^{-1/3}.
\end{align}
A nearby TLF will generate a small fluctuation $dF_z =  z_d/|\mathbf{r}-\mathbf{r}_{d}|^3$ (as the qubit is in the $z=0$ plane) away from $F_z$ and lead to a variation of the $g$-factor by,
\begin{align}
    dg(\phi, F_z) = \frac{\partial g}{\partial F_z}(\phi, F_z) dF_z = 1.5587\frac{2|e|^{2/3}}{3\mu_B}\left( \frac{1}{2m_l\hbar}\right)^{1/3}\left( \beta^{'}\sin(2\phi)-\alpha^{'}\right) F_z^{-1/3} \frac{e}{4\pi \varepsilon}\frac{z_d}{|\mathbf{r}-\mathbf{r}_{d}|^3} 
    = \sigma(\phi, F_z)\frac{z_d}{|\mathbf{r}-\mathbf{r}_{d}|^3}.
\end{align}
where we define $\sigma(\phi, F_z) = 1.5587\frac{e^{5/3}}{6\pi\mu_B\varepsilon}\left( \frac{1}{2m_l\hbar}\right)^{1/3}\left( \beta^{'}\sin(2\phi)-\alpha^{'}\right) F_z^{-1/3}$. In the main text, we combined $g_0$ and $\delta g(\phi, F_z)$ to a single term $g_0$, as both terms remain constant for a flat interface and fixed magnetic field angle.
Realistic values for the different parameters are given in \cref{tab:g_factor_params}, leading to $\sigma^{'} =  \mu_B \sigma B_0/\hbar \approx 3186$ nm$^{2}/\mu$s.
\begin{table}[h!]
    \centering
    \begin{tabular}{l|c}
        \hline
        \hline
         \quad \quad  \quad \quad \quad Parameters & Value \\
         \hline
         Rashba coupling (estimated based on \cite{ferdous_interface_2018}) & $2\ \mu \text{eV} \cdot \text{nm}$  \\
         Dresselhaus coupling (estimated based on \cite{ferdous_interface_2018}) &  $200\ \mu \text{eV} \cdot \text{nm}$ \\
         Longitudinal effective mass $m_l$ \cite{dexter1954effective} &  $0.98 m_0$ \\
         Confinement electric field $F_z$ (chosen based on \cite{ferdous_interface_2018})  & $30\ \text{MV/m}$\\
         Magnetic field angle $\phi$ & $\pi /4$ \\
         Magnetic field $B_0$ & 1 T \\
         Relative electric permittivity $\varepsilon/\varepsilon_0$ & 7.8
         \\
         \hline
         \hline
    \end{tabular}
    \caption{Parameters used to derive the proportionality constant between the $g$-factor of a qubit and the electric field generated by a nearby TLF.}
    \label{tab:g_factor_params}
\end{table}
\section{\label{app:proofs} Formal derivations}
\subsection{Derivation of the dephasing factor}
In this section, we derive an analytical model for $W(T)$.
Suppose that we start with the TLF in $\mathbf{r}_{d}^{+}$ at time $t$ and consider what could happen during the interval $[t, t+dt]$. Let $\gamma$ be the switching rate; we have the following two scenarios:
Either no jump occurred in the interval, which happens with probability $\exp(-\gamma dt)$, 
or a jump occurred at time $t+s$ with $0 \leq s \leq dt$ which happens with probability $1-\exp(-\gamma dt)$.
We thus have the following decomposition,
\begin{align}
    w_+(t+dt) = \exp(-\gamma dt)\exp(-if_+(t)dt)w_+(t) + (1-\exp(-\gamma dt))\exp(-if_-(t)s)\exp(-if_+(t)(dt-s))w_-(t).
\end{align}
A similar decomposition can be made for $w_-(t+dt)$. 
Consider the first derivative of $w_\pm$,
\begin{align}
    \frac{dw_\pm}{dt}(t) = \lim_{dt\rightarrow0} \frac{w_\pm(t+dt)-w_\pm(t)}{dt}.
\end{align}
To compute it, we only need to obtain a decomposition of $w_\pm(t+dt)$ to first order in  $dt$. We have, e.g.,
\begin{align}
    w_{+}(t+dt) &\approx (1-\gamma dt) (1-if_+(t)dt)w_+(t) + \gamma dt (1-if_-(t)s)(1-if_+(t)(dt-s))w_-(t) \skipLine 
    &\approx w_+(t) + \left(-if_+(t)-\gamma\right)w_+(t)dt+ \gamma w_-(t)dt,
\end{align}
where we expanded each exponential to first order in the first line and kept only the first order terms (in $dt$) in the second line.
This yields,
\begin{align}
    \frac{dw_+}{dt}(t) = (-if_+(t)-\gamma)w_+(t)+\gamma w_-(t),
\end{align}
and similarly,
\begin{align}
    \frac{dw_-}{dt}(t) = (-if_-(t)-\gamma)w_-(t)+\gamma w_+(t).
\end{align}
Combining these two equations we find the system of differential equations,
\begin{align}
    \frac{d\mathbf{w}}{dt}(t) = M(t)\mathbf{w}(t),
\end{align}
with
$M(t) = \begin{pmatrix}
    -if_+(t)-\gamma & \gamma \\
    \gamma & -if_-(t)-\gamma
\end{pmatrix}$ and $f_\pm(t) = \sigma^{'}z_d^{\pm}/|\mathbf{r}(t)-\mathbf{r}_{d}^\pm|^3$.

\subsection{The $\gamma T \ll 1$ limit}
When $\gamma T \ll 1$, the system of equations can be rewritten as two independent differential equations,
\begin{align}
    \frac{dw_+}{dt}(t) = -if_+(t)w_+(t), \skipLine
    \frac{dw_-}{dt}(t) = -if_-(t)w_-(t),
\end{align}
which gives us 
\begin{align}
    w_+(T) = w_+(0)\exp\left(-i\int_0^Tf_+(\tau)d\tau\right), \skipLine
    w_-(T) = w_-(0)\exp\left(-i\int_0^Tf_-(\tau)d\tau\right).
\end{align}
More precisely, 
\begin{align}
    \int_0^Tf_\pm(\tau)d\tau &= \sigma^{'}\Bigg(\int_0^{T/2}\frac{z_0\pm z_1}{\left((v\tau-x_0\mp x_1)^2 + (y_0 \pm y_1)^2 + (z_0 \pm z_1)^2\right)^{3/2}}d\tau \skipLine
    &+ \int_{T/2}^{T}\frac{z_0 \pm z_1}{\left((2L-v\tau-x_0 \mp x_1)^2 + (y_0 \pm y_1)^2 + (z_0 \pm z_1)^2\right)^{3/2}}d\tau\Bigg).
\end{align}
By performing the change of variable $s_\pm = x_0 \pm x_1 - v\tau$ (and $s_\pm = x_0 \pm x_1 - L + v\tau$ for the second integral) we get,
\begin{align}
    &\int_0^Tf_\pm(\tau)d\tau = \frac{2\sigma^{'}(z_0\pm z_1)}{v}\int_{x_0 \pm x_1-L}^{x_0\pm x_1}\frac{ds}{\left(s_\pm^2 + (y_0 \pm y_1)^2 + (z_0 \pm z_1)^2\right)^{3/2}},
\end{align}
where we used $L = vT/2$.
The factor $2$ comes from the fact that the two integrals are equal. This is a known integral and we can get the following analytical form,
\begin{align}
    \int_0^T f_\pm(\tau)d\tau = \frac{2\sigma^{'}(z_0 \pm z_1)}{v\beta_\pm^2}\left( \frac{L-(x_0 \pm x_1)}{\sqrt{(L-(x_0\pm x_1))^2+\beta_\pm^2}} + \frac{x_0 \pm x_1}{\sqrt{(x_0\pm x_1)^2+\beta_\pm^2}}\right),
\end{align}
with $\beta_\pm = \sqrt{(y_0\pm y_1)^2 + (z_0\pm z_1)^2}$.

\subsection{The $\gamma T \gg 1$ limit}
In this regime, we have, 
\begin{align}
    \frac{du_1}{dt}(t) &= -i \bar{f}(t)u_1(t) -i\delta f(t)u_2(t), \skipLine
    \frac{du_2}{dt}(t) &= -i \bar{f}(t)u_2(t) -i\delta f(t)u_1(t) - 2\gamma u_2(t),
    \label{eq:gamma_inf_sys}
\end{align}
where $u_1 = w_+ + w_-$, $u_2 = w_+-w_-$, $\bar{f}(t) = 1/2(f_+(t)+f_-(t))$ and $\delta f(t) = 1/2(f_+(t)-f_-(t))$.
As $\gamma T \gg 1$, we can see that $u_2$ will decay extremely fast, we can thus assume that $\frac{du_2}{dt} = 0$. This leads to,
\begin{align}
    u_2(t) &= -\frac{i\delta f(t)}{i\bar{f}(t)+2\gamma}u_1(t) \skipLine
    &= -\left( \frac{\delta f(t) \bar{f}(t)}{4\gamma^2 + \bar{f}(t)^2} +i\frac{2\gamma \delta f(t)}{4\gamma^2 + \bar{f}(t)^2}\right) u_1(t) \skipLine
    &\approx -\left(\frac{\delta f(t) \bar{f}(t)}{4\gamma^2} +i\frac{\delta f(t)}{2\gamma}\right)u_1(t),
\end{align}
where we used the fact that $\gamma \gg |\bar f(t)|$ in the last step.
Substituting this into the first equation in \cref{eq:gamma_inf_sys}, we find,
\begin{align}
    \frac{du_1}{dt}(t) &= \left( -\frac{\delta f(t)^2}{2\gamma} - i\left(1 - \frac{\delta f(t)^2}{4\gamma^2} \right)\bar{f}(t)\right)u_1(t) \skipLine
    &\approx \left( -\frac{\delta f(t)^2}{2\gamma} - i\bar{f}(t)\right)u_1(t),
\end{align}
where we used the fact that $\gamma \gg |\delta f(t)|$ in the last step.
This then gives us
\begin{align}
    W(T) = u_1(T) = \exp\left( -i\int_{0}^{T}\bar{f}(\tau)d\tau\right)\exp\left( -\frac{1}{2\gamma}\int_{0}^{T}\delta{f}(\tau)^2d\tau\right),
\end{align}
where we assumed that $u_1(0) = w_+(0)+w_-(0) = 1$.

As we are only concerned with computing the infidelity $1-|W(T)|$, we can focus on the analytic form of $\exp\left( -\frac{1}{2\gamma}\int_0^T\delta f(t)^2dt\right)$. For simplicity, we do not include this lengthy expression here, and rather use a numerical solver to compute it.

\section{\label{app:alt_protocols} Alternative protocols}
In the main text, we propose to detect the position of a TLF by shuttling an electron back and forth for various distances (FB protocol). However, this is not the only way to locate the charge defect. Any protocol allowing for a different phase to be picked up depending on the location of the qubit will work. 

\begin{figure}
    \centering
    \includegraphics[width=\linewidth]{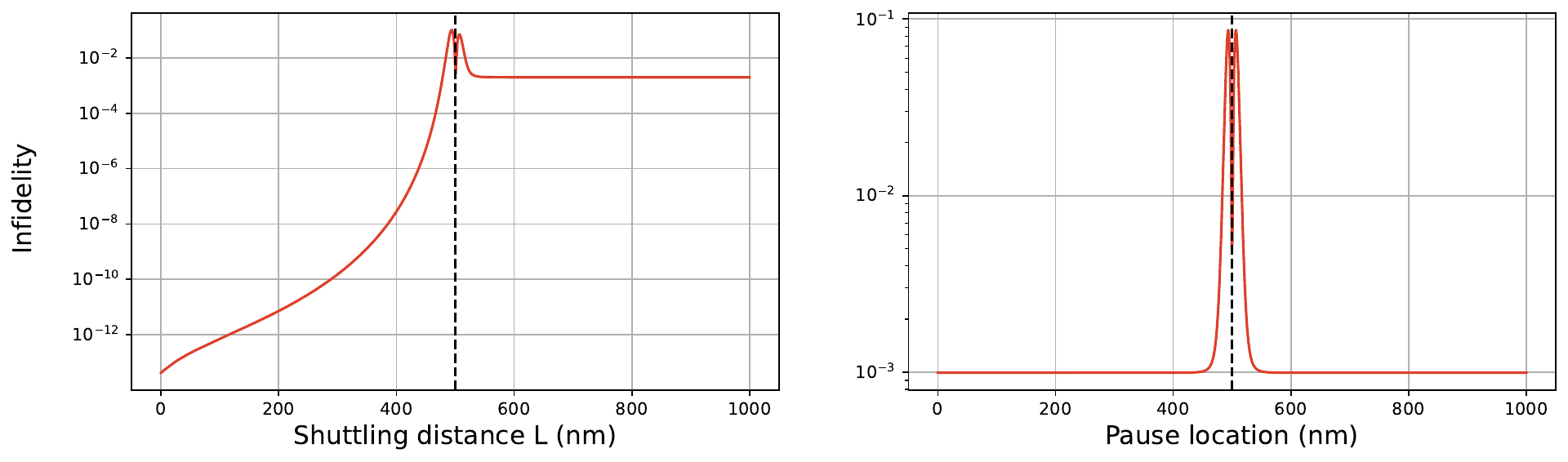}
    \caption{Infidelity with respect to shuttling distance when the TLF is fluctuating in the $x$-direction. The left panel corresponds to an electron being shuttled back and forth with a $1 \;\mu$s pause at the turning point (FWB protocol). In the right panel, we study an electron travelling in one direction from $x= 0$ to $x= 1000$ nm. The x axis corresponds to the position at which the electron is stopped for $1 \;\mu$s to accumulate phase before resuming its journey (FWF protocol). The mean position of the defect is chosen to be $x_0=500\,{\rm nm}$ (vertical dashed line). Generally,  parameters are chosen according to Table~\ref{tab:parameters}. }
    \label{fig:alt_protocol_x}
\end{figure}

Here, we propose two alternative protocols. In the forward-wait-backward (FWB) protocol, an electron is being shuttled back and forth as in the main text, but a pause (waiting time) is added at the turning point to increase the impact of the TLF. In the forward-wait-forward (FWF) protocol, the electron is shuttled only one way and paused at various locations. The choice of which protocol to implement will depend on the available experimental setup. 
In \cref{fig:alt_protocol_x}, we plot the evolution of the infidelity for both protocols for a TLF oscillating in the $x$-direction for a pause time of $1 \;\mu$s. 
Although the infidelity near the defect is higher for these protocols compared to the one presented in the main text, we decided to present the protocol without pause in the main text as it is the simplest to run and interpret. 

\section{\label{app:params} Model Parameters}

The parameters chosen for the simulation were based on RF reflectometry measurements performed by the Quantum Hardware team of Quantum Motion. The trace of the charging voltage was analysed with a Factorial Hidden Markov Model (FHMM) \cite{ghahramani1995factorial}. Hidden Markov Models use the expectation maximization algorithm \cite{em_algorithm} to estimate the hidden states of a Markov process based on some observation, extracting transition probabilities and amplitudes. A \emph{factorial} HMM generalises this further, and instead of looking at a dense transition matrix between all possible hidden states, it considers every hidden state as a sum of multiple underlying Markov processes. This means that, e.g., for 3 TLFs, an HMM would need $2^3$ hidden states and therefore fit an $8\times8$ dense transition matrix, whereas an FHMM would use three separate $2\times2$ transition matrices instead. This is computationally more efficient and better represents the underlying physics, giving us direct access to the parameters of individual TLFs.
\section{\label{app:other_dir} Charge fluctuating in other directions}
\begin{figure}
    \centering
    \includegraphics[width=\linewidth]{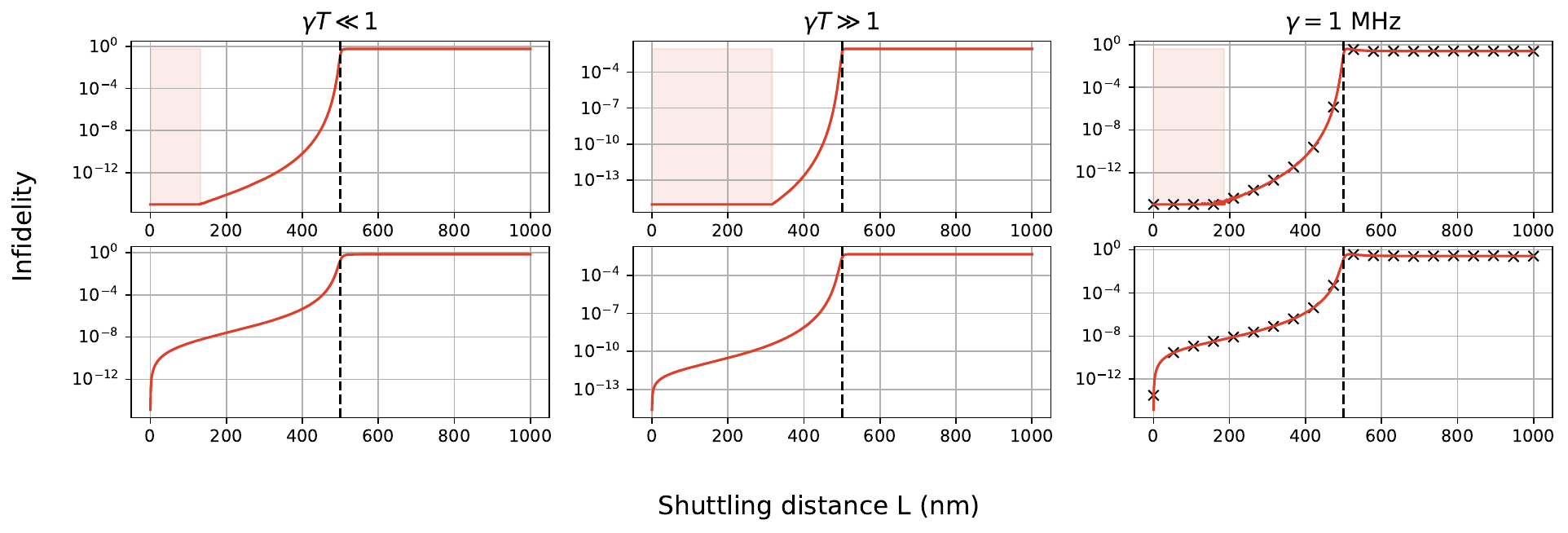}
    \caption{Infidelity with respect to shuttling distance when the TLF is fluctuating in the $y$-direction (top) and $z$-direction (bottom) for $\gamma T \ll 1$ (left), $\gamma T \gg 1$ (middle) and $\gamma = 1 $ MHz (right). The mean position of the defect is chosen to be $x_0=500\,{\rm nm}$ (vertical dashed line). Generally,  parameters are chosen according to Table~\ref{tab:parameters}. Due to numerical instability and precision constraints, values within the red shaded region have been clipped.}
    \label{fig:other_directions}
\end{figure}

In \cref{fig:other_directions} we plot the evolution of the infidelity for TLF fluctuating in $y$-direction (top) and $z$-direction (bottom). As the effect is less strong than in the $x$-direction, we sometime reach machine precision far from the defect. To remove the numerical instability in these regions (red shaded areas), we clipped the infidelity to $10^{-14}$.

\section{\label{app:more_defects} Detecting the position of more than one defect}

\begin{figure}
    \centering
    \includegraphics[width=1\linewidth]{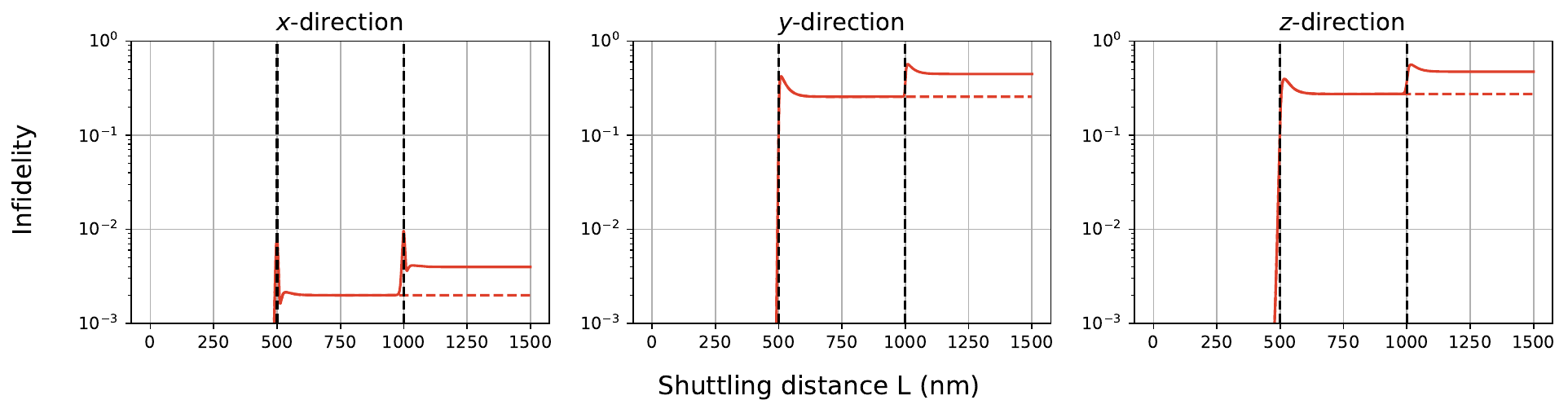}
    \caption{Infidelity versus shuttling distance $L$ in the $\gamma = 1$~MHz regime for one (dashed red) and two defects (solid red). The two vertical dashed lines represent the defects' mean positions: $x_{0,1}=500$~nm and $x_{0,2}=1000$~nm. All other parameters follow Table~\ref{tab:parameters}.}
    \label{fig:two_defects}
\end{figure}

In the main text, we focus on detecting a single charge defect, but our protocol is not fundamentally limited to this case. To demonstrate this, we numerically solve a generalised version of the coupled differential
equations in \cref{eq:sys_diff_eq} for two TLFs, placing the first defect at $x_{0,1} = 500$~nm and the second at $x_{0,2} = 1000$~nm. All
other parameters held at the values in Table~\ref{tab:parameters}. \cref{fig:two_defects}, which zooms on the region of interest, shows that both defects remain detectable in the realistic $\gamma = 1$~MHz regime. Our protocol therefore generalises naturally to multiple defects, provided they are sufficiently well separated.

\section{\label{app:spin} A Note on Spin}

An important advantage of our method is that it is sensitive to charge traps in their \emph{neutral} state, through the interaction of the spin of the unpaired electron in a $P_b$ centre and the spin of the qubit electron. (Note that the spin-interactions of $P_b$ centres have been studied before \cite{Poindexter1981InterfaceSA}.) Considering the long activation-passivation cycles \cite{defect_candidates}, it is a concern that such a neutral trap might stay hidden for a long time, then activate after calibration has been done. It is important to point out that the spin-spin effect is expected to be much smaller than the electrostatic interaction, but considering the advantage it yields, its investigation is warranted.

\end{document}